\documentclass[aps,prd,showpacs,preprintnumbers,twelvepoint,floatfix]{revtex4}
\usepackage{epsfig}
\usepackage{latexsym,amsmath,amssymb,amstext}
\usepackage{float}
\usepackage{graphicx}
\usepackage{color}

\newcommand\bef{\begin{figure}}
\newcommand\eef[1]{\label{fg:#1}\end{figure}}
\newcommand\beq{\begin{equation}}
\newcommand\eeq[1]{\label{#1}\end{equation}}
\newcommand\beqa{\begin{eqnarray}}
\newcommand\eeqa[1]{\label{#1}\end{eqnarray}}
\newcommand\bet{\begin{table}}
\newcommand\eet[1]{\label{tb:#1}\end{table}}

\newcommand\fgn[1]{Figure \ref{fg:#1}}
\newcommand\eqn[1]{eq.\ (\ref{#1})}
\newcommand\scn[1]{Section \ref{sec:#1}}

\newcommand\tbn[1]{Table \ref{tb:#1}}

\newcommand\ie{{\sl i.e.\/}}

\newcommand\etal{{\sl et al.\/}}
\newcommand\jhep{{\sl J.\ H.\ E.\ P.\/}\ }
\newcommand\np{{\sl Nucl.\ Phys.\/}\ }

\newcommand{\MSbar}{{\overline{\scriptscriptstyle MS}}}
\newcommand{\tr}{{\mathrm{tr}\/}}
\newcommand\rval{1.106 \pm 0.007 ({\rm stat\/}) \pm 0.005 ({\rm syst})}

\begin{document}
\title{Wilson flow with naive staggered quarks}
\author{Saumen\ \surname{Datta}}
\email{saumen@theory.tifr.res.in}
\affiliation{Department of Theoretical Physics, Tata Institute of Fundamental
         Research,\\ Homi Bhabha Road, Mumbai 400005, India.}
\author{Sourendu\ \surname{Gupta}}
\email{sgupta@theory.tifr.res.in}
\affiliation{Department of Theoretical Physics, Tata Institute of Fundamental
         Research,\\ Homi Bhabha Road, Mumbai 400005, India.}
\author{Anirban\ \surname{Lahiri}}
\email{anirban@theory.tifr.res.in}
\affiliation{Department of Theoretical Physics, Tata Institute of Fundamental
         Research,\\ Homi Bhabha Road, Mumbai 400005, India.}
\author{Andrew\ \surname{Lytle}}
\email{andrew.lytle@glasgow.ac.uk}
\affiliation{SUPA, School of Physics and Astronomy, University of Glasgow,
         Glasgow G12 8QQ, UK.}
\author{Pushan\ \surname{Majumdar}}
\email{tppm@iacs.res.in}
\affiliation{Department of Theoretical Physics, Indian Association for the
         Cultivation of Science,\\ Raja Subodh Chandra Mallick Road,
         Jadavpur, Kolkata 700032, India.}

\begin{abstract}
Scale setting for QCD with two flavours of staggered quarks is examined
using Wilson flow over a factor of four change in both the lattice spacing
and the pion mass.  The statistics needed to keep the errors in the flow
scale fixed is found to increase approximately as the inverse square
of the lattice spacing. Tree level improvement of the scales $t_0$ and
$w_0$ is found to be useful in most of the range of lattice spacings we
explore. The scale uncertainty due to remaining lattice spacing effects
is found to be about 3\%. The ratio $w_0/\sqrt{t_0}$ is $N_f$ dependent
and we find its continuum limit to be $\rval$ for $m_\pi w_0\simeq0.3$.
\end{abstract}

\preprint{TIFR/TH/15-20}
\maketitle

\section{Introduction}\label{sec:intro}

In any cutoff field theory it is easy to set the unit of mass in terms of
the momentum cutoff. So, in lattice field theories the scale can be set
by the inverse lattice spacing, $1/a$. However, physically interesting
questions require us to relate one measurable quantity to another,
when both are computed to comparable precision in the theory. Using a
physical scale to set the units of mass by eliminating the artificial
choice of $a$ is called setting the lattice scale.  Doing this allows us
to take the limit $a\to0$ in renormalizable theories without encountering
artificial infinities.

In principle, any mass scale can be chosen to define units, so
the question of what to use for a mass scale is essentially one of
convenience. An ideal scale should be easy to control numerically in the
non-perturbative domain as well as be amenable to perturbative analysis.
In recent years Wilson flow \cite{flow,also} has emerged as a new and
computationally cheap way of setting the lattice scale \cite{bmw,sommer},
since it seems to fulfill both criteria.

However, Wilson flow scales, like $\Lambda_\MSbar$, are theory scales.
In order to determine them in ``physical'' (GeV) units, one needs two
separate scale computations within the theory: one of the theory scale
under question, the other of a measurable scale. Then by comparing the
measurable scale to experiment, one can determine the theory scale in
physical units. Clearly, in order to do this one needs to control two
measurements. For Wilson flow this has been attempted in quenched QCD
\cite{flowqcdscale}, with 2 flavours of Wilson quarks \cite{wilson2},
2+1 flavours of improved Wilson \cite{wilson21,bmw} and improved
staggered quarks \cite{bmw} and 2+1+1 flavours of improved staggered
quarks \cite{milcscale}.

These computations have uncovered several systematics in the setting of
the Wilson scale. In this paper we investigate in detail these systematics
for two flavours of naive staggered quarks over a large range of lattice
spacing and pion mass. We report on investigations of statistical
uncertainties, as well as the dependence on all tunable parameters. We
present an estimate of the Wilson flow scale in physical units.

In the next section we outline the methods which we use. In \scn{runs}
we present a summary of the runs and statistics. A description of our
results is given in \scn{results}, and a summary given in \scn{conclude}.

\section{Methods and Definitions}\label{sec:method}

Start with a gauge field configuration, \ie, the set of link matrices,
$\{U_\mu(x)\}$, where $x$ denotes a point in the 4-d Euclidean
space-time lattice, and $\mu$ denotes one of the 4 directions. Wilson
flow of this configuration is the evolution of these matrices in a
fictitious ``flow time'' $t$, using the differential equation
\beq
   \frac{dU_\mu(x,t)}{dt} = -\frac{\partial S[U]}{\partial U_\mu(x)} U_\mu(x,t),
     \quad{\rm where}\quad U_\mu(x,0)=U_\mu(x),
\eeq{floweq}
and the derivative is the usual Hermitean traceless matrix obtained by
differentiating the scalar valued action functional $S[U]$ with respect
to the link matrix \cite{montvay}. We use the convention that
\beq
   S[U]=\sum{\rm Re\/}\,\tr\,[1-U(p)],
\eeq{action}
where $U(p)$ is the ordered product of link matrices around a plaquette,
and the sum is over plaquettes. Clearly, the configuration with all $U=1$
is a fixed point of the flow, and it can be shown that it is an attractive
fixed point with a finite basin of attraction \cite{flow}.

Following \cite{flow}, we define the scale by constructing the quantity
\beq
   {\cal E}(t)=t^2 E(t), \qquad{\rm where}\qquad
   E(t) = -\frac12\overline{\tr F_{\mu\nu}(x,t)F^{\mu\nu}(x,t)},
\eeq{energy}
where $F_{\mu\nu}$ is a lattice approximation to the gluon field strength
tensor and the bar denotes averaging over the lattice volume. The field
strength tensor can be built either from the Wilson plaquette operator
or through a 16-link clover operator. Some of our investigation of the
systematics of Wilson flow involves comparing these two definitions. The
scales which emerge from this are defined through the equations
\beq
  \left.\langle{\cal E}(t)\rangle\right|_{t=t_0(c)} = c, \qquad
  \left.t \frac{d\langle{\cal E}(t)\rangle}{dt}\right|_{t=w_0^2(c)}=c.
\eeq{flowtimes}
The choice of $c=0.3$ gives the quantities usually referred to as $t_0$
and $w_0$ in the literature, a convention that we adopt. The
modification, $c=2/3$ has also been suggested \cite{sommer}. The value
$c=0.4$ has been used in \cite{flowqcdscale}. A weak coupling expansion
\cite{flow} gives
\beq
   \langle{\cal E}(t)\rangle = \frac3{(4\pi)^2} g^2 + {\cal O}(g^4).
\eeq{weakcoupling}
If one uses $t_0(c)$ to set the scale, then the expression above can be
used to define a renormalized coupling
\beq
   g_R^2 = \frac{16\pi^2c}3,
\eeq{coupling}
and it is clear that the choice of $c$ is equivalent to a choice of
the renormalization scheme. We report a study of this choice later in
this paper. Note that the values of $c$ used generally correspond to
$\alpha_S=g_R^2/(4\pi)>1$.

Tree-level improvement was performed by noting that the weak-coupling
expansion in \eqn{weakcoupling} can be systematically corrected for
lattice-spacing dependence through a computable piece
\beq
   \langle{\cal E}(t)\rangle = \frac3{(4\pi)^2} g^2 C\left(\frac{a^2}t\right)
     \quad{\rm where}\quad
   C\left(\frac{a^2}t\right) = 1 + \sum_{m=1}^\infty C_{2m}
         \left(\frac{a^2}t\right)^m.
\eeq{treelevel}
We use the coefficients presented in \cite{tlimprov}.  Later in this
paper we show the effect of these corrections, and incorporate them in
our measurements of the scale.

We have also incorporated a finite volume correction due to the zero-mode
of the gauge field \cite{volzero}. Its effect is to scale
\beq
   c \to c \left[1-\frac{\zeta^4\pi^2}3+\vartheta\left(
       {\rm e}^{-1/\zeta^2}\right)\right] \approx
   c \left[1-\frac{\zeta^4\pi^2}3+8{\rm e}^{-1/\zeta^2}\left(
               1+3{\rm e}^{-1/\zeta^2}\right)\right]
\eeq{volzero}
where $\zeta=\sqrt{8t}/L$, $L$ is the lattice extent, and $\vartheta$
is a Jacobi Theta function. Except at our two smallest bare couplings,
the effect of the finite volume correction is comparable to, or smaller
than, the statistical errors.

\section{Runs}\label{sec:runs}

\bet
\begin{tabular}{l|l|c||c|c||c||c|c}
\hline
$\beta$ & $ma$ & $N_s$ & Machine & Traj & Statistics 
 & $w_0/a$ & $m_\pi a$ \\
 & & $L/a$ & & (MD) & $T_0+T \times N$ & & \\
\hline
5.2875	& 0.1	& 16 & V & 1 & $400+10\times50$ & 0.6112 (4) & 0.790 (1) \\
	& 0.05	& 16 & V & 1 & $780+10\times50$ & 0.6354 (6) & 0.575 (1) \\
	& 0.025	& 16 & V & 1 & $200+15\times70$ & 0.6539 (1) & 0.415 (2) \\
	& 0.015	& 16 & V & 1 & $400+10\times50$ & 0.6608 (5) & 0.325 (2) \\
\hline
5.4	& 0.05	& 16 & V & 2 & $200+20\times75$ & 0.8418 (14) & 0.604 (2) \\
	& 0.025 & 16 & V & 1 & $400+10\times51$ & 0.9264 (21) & 0.443 (2) \\
	& 0.015	& 24 & V & 2 & $400+10\times50$ & 0.9600 (9)  & 0.351 (1) \\
	& 0.01	& 32 & G & 2 & $200+20\times40$ & 0.9922 (7)  & 0.292 (2) \\
\hline
5.5	& 0.05	& 16 & V & 1 & $200+20\times50$   & 1.1689 (40) & 0.613 (2) \\
	& 0.025	& 24 & V & 1 & $1680+10\times101$ & 1.2651 (18) & 0.446 (1) \\
	& 0.015	& 28 & G & 2 & $400+10\times120$  & 1.3302 (13) & 0.353 (2) \\
	& 0.01	& 32 & G & 2 & $200+20\times40$   & 1.3771 (16) & 0.294 (2) \\
	& 0.005	& 32 & BG & 1 & $250+10\times50$  & 1.4254 (37) & 0.212 (1) \\
\hline
5.6	& 0.05	& 24 & V & 1 & $400+10\times55$  & 1.4850 (26) & 0.594 (2) \\
	& 0.025	& 24 & V & 1 & $1700+10\times48$ & 1.6007 (33) & 0.427 (2) \\
	& 0.015	& 28 & G & 2 & $400+10\times120$ & 1.7087 (25) & 0.329 (2) \\
	& 0.01	& 32 & G & 2 & $200+20\times40$  & 1.7814 (36) & 0.272 (2) \\
	& 0.005	& 32 & BG & 1 & $300+10\times50$ & 1.8547 (71) & 0.198 (2) \\
	& 0.003	& 32 & BG & 1 & $600+5\times105$ & 1.8824 (32) & 0.151 (1) \\
\hline
5.7	& 0.025	& 24 & V & 1 & $530+10\times59$  & 1.9645 (48) & 0.395 (2) \\
	& 0.005	& 32 & BG & 1 & $370+10\times50$ & 2.1470 (73) & 0.177 (3) \\
	& 0.003	& 32 & BG & 1 & $300+10\times50$ & 2.2103 (162) & 0.139 (7) \\
	& 0.002	& 32 & BG & 1 & $480+5\times62$  & 2.3765 (67) & --- \\
\hline
\end{tabular}
\caption{The data sets used in this paper. Runs were made at different bare
 couplings $\beta$ and bare quark masses $ma$. Hypercubic $N_s^4$ lattices
 were used, where $N_s=L/a$. $N$ gauge configurations were collected for
 each run, after discarding an initial time $T_0$ for thermalization, and
 collecting one configuration after every time $T$ (both $T_0$ and $T$ are
 given in MD time units). The runs were performed on a vector machine (V),
 on a Blue Gene (BG) and GPUs (G). For completeness, our estimates of the
 tree-level improved flow scale $w_0/a$ and the pseudo-Goldstone pion mass
 $m_\pi a$ are also collected here, although they are discussed in detail
 in \scn{results}.}
\eet{configs}

We generated gauge field configurations with two flavours of
naive staggered quarks over a wide range of bare couplings and
bare quark masses. The bare parameters and statistics are given in
\tbn{configs}. Since the runs were performed on different machines we
took the precaution of repeating several runs on multiple machines in order
to cross check results. In these cases only the runs with the largest
statistics are reported in the table above.

A part of this range has been explored earlier, and pion masses have
been reported \cite{nf2}.  We checked that at the common points our
measurements of pion masses agree with those previously reported
in the literature. All our analyses use the bootstrap technique
to estimate expectation values and errors. Since we estimate pion
correlation functions using a bootstrap, we reduce covariances between
measurements at different distances by using independent bootstraps
at each distance. Confidence intervals on the fitted parameters are
estimated by a bootstrap over fits.

The biggest challenge in estimating pion masses at small lattice spacings
is in using lattice extents which are large enough to separate out the
ground state from excitations. This is most acute for the Goldstone
pseudoscalar mass at the smallest bare quark mass and lattice spacing,
where our lattice size ($m_\pi L<4$) was clearly inadequate.  So we
do not quote the pion mass from this lattice.  Since we use naive
staggered quarks, taste symmetry breaking remains a concern. We will
report investigations of this elsewhere.

\bef
\includegraphics[scale=1.0]{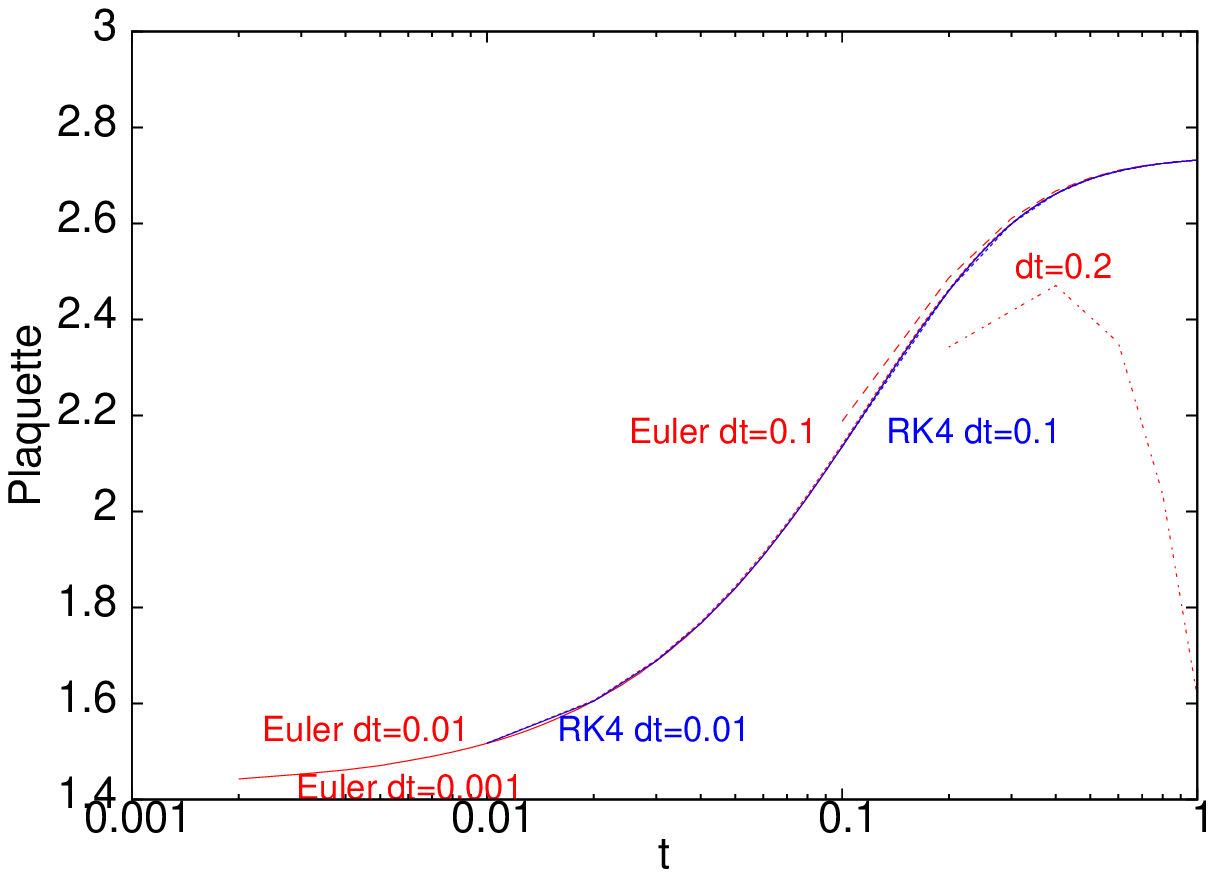}
\caption{The effect on the flow of changing integrator and $dt$. For
 $dt<0.2$ the difference between integrators is negligible, especially
 with increasing $t$. For $dt\ge0.2$ the integrator wanders off from
 the true solution.}
\eef{integ}

\bef
\includegraphics[scale=1.0]{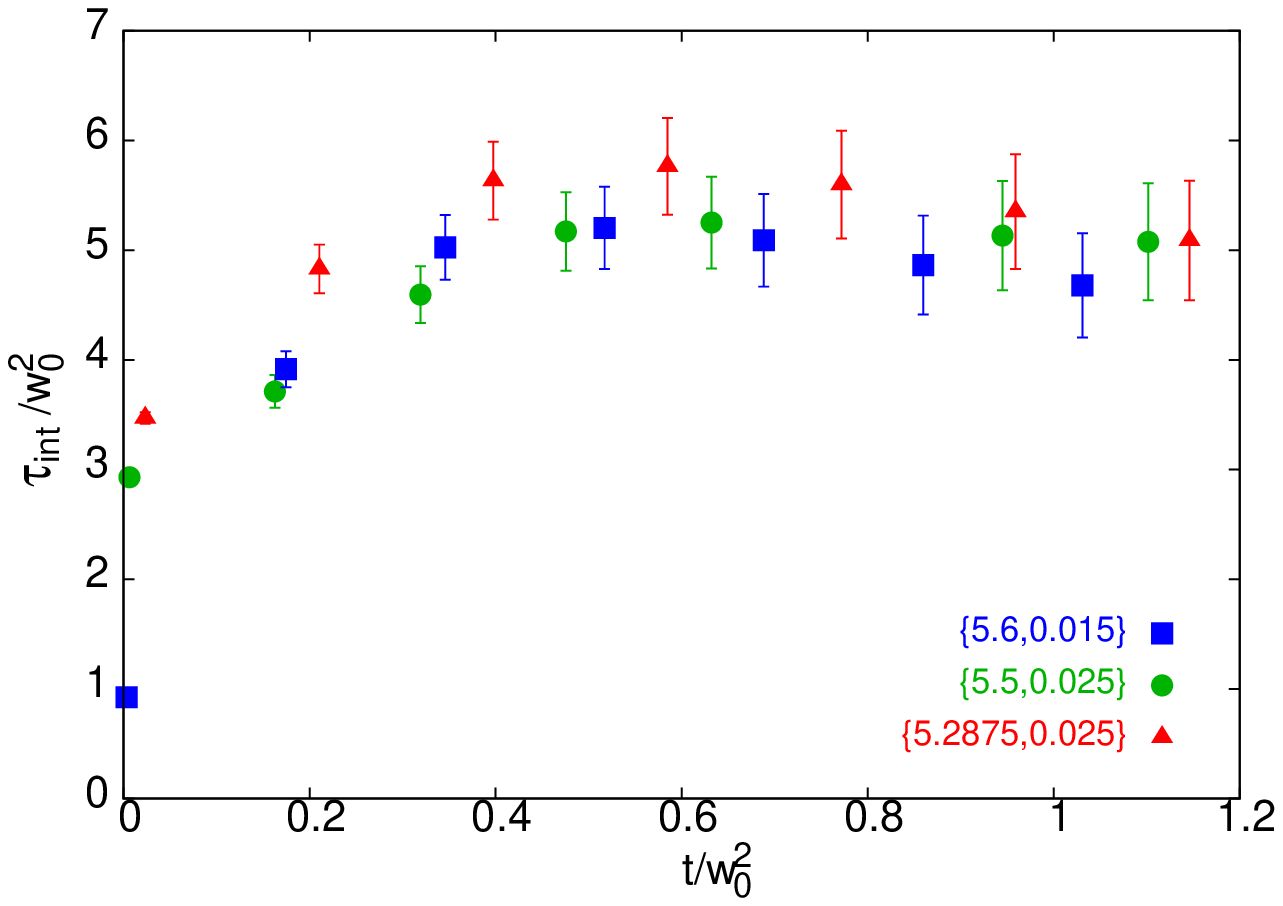}
\caption{The integrated autocorrelation time estimated for ${\cal E}(t)$
 computed from the Wilson operator for several data sets. The autocorrelations
 increase very rapidly with flow time before running out of statistics.}
\eef{tauint}

One technical issue has to do with the integration of the flow
equations. We tested both the Euler integrator and the fourth-order
Runge Kutta (RK4) integrator. In \fgn{integ} we show the evolution of
the plaquette under the flow when it is integrated using each of these
methods for one fixed configuration. As shown, both integrators perform
well even for $dt=0.1$. Note that with the Euler integrator and $dt=0.1$
the first integration step has larger errors than the later steps. Such
self-repair is seen also with a finer time step of $dt=0.01$. With the
Euler integrator one sees a failure of this self-repairing mechanism when
$dt=0.2$. The observation that the free configuration is an
attractor of the map in \eqn{floweq} serves to explain both self-repair
and its failure. The fact that there is an attractor with a finite basin
of attraction is the reason for self-repair, with the global errors
being smaller than the ${\cal O}(t^2)$ predicted by a local analysis.
Its failure occurs for sufficiently coarse $dt$, when the flow falls
outside this basin of attraction. RK4 is generally more stable, but
even so, its global error is smaller than local analysis would lead
us to believe.  We used RK4 with $dt=0.01$, but checked the
results statistically by changing $dt$ by a factor of 4 either way. We
found that the statistical uncertainty in the measurement of flow times
is larger than any effect of the evolution.

The statistical properties of the measurement of ${\cal E}(t)$ under
evolution in flow time are also of interest. Since our measurements
are separated by 10 or 20 MD trajectories, at $t=0$ they are quite
decorrelated. However, as the flow integrates information over
successively larger volumes, one expects autocorrelations to grow
with flow time. We quantify the autocorrelations in terms of the integrated
autocorrelation time, $\tau_{\rm int\/}$, which is defined in terms of
an autocorrelation function of the measurements $C(s)$ as
\beq
   \tau_{\rm int\/} = 1 + 2 \int_0^\infty ds\, C(s),
\eeq{tauint}
where $s$ is the separation between the measurements.
In \fgn{tauint} we show 
$\tau_{\rm int\/}$ for ${\cal E}(t)$ as a function
of the flow time, $t$. In accordance with expectations, this shows an
initial rapid increase.  The observed plateau in $\tau_{\rm int\/}$ is
due to insufficient statistics; clearly for the set with $\beta=5.6$,
the effective number of configurations decreases by a factor of around
20 when this plateau develops.  Significantly more statistics would
be needed to improve the measurement in this region, and decrease the
estimate of the error in $\tau_{\rm int\/}$ elsewhere.

If the scaling of autocorrelations is physical, \ie, has a sensible
continuum limit, then a natural way to compare flow times for different
simulations would be to scale them by $w_0^2$ (or, equivalently,
$t_0$). HMC simulations with fixed trajectory lengths have $\tau_{\rm
int\/}$ scaling as the square of correlation lengths \cite{hmc}.
Since flow time also scales as the square of lengths, one should expect
$\tau_{\rm int\/}\simeq w_0^2$. \fgn{tauint} illustrates that this scaling
is not present in the initial state, but develops fairly early during the
flow and is a good first approximation to the observation.  It would take
significantly improved statistics to study the remaining deviations. The
physics result is simple: as the lattice spacing decreases, the statistics
required to keep a constant error on the flow scale increases (roughly)
as the inverse square of the lattice spacing.

\section{Results}\label{sec:results}

\bef
\includegraphics[scale=0.7]{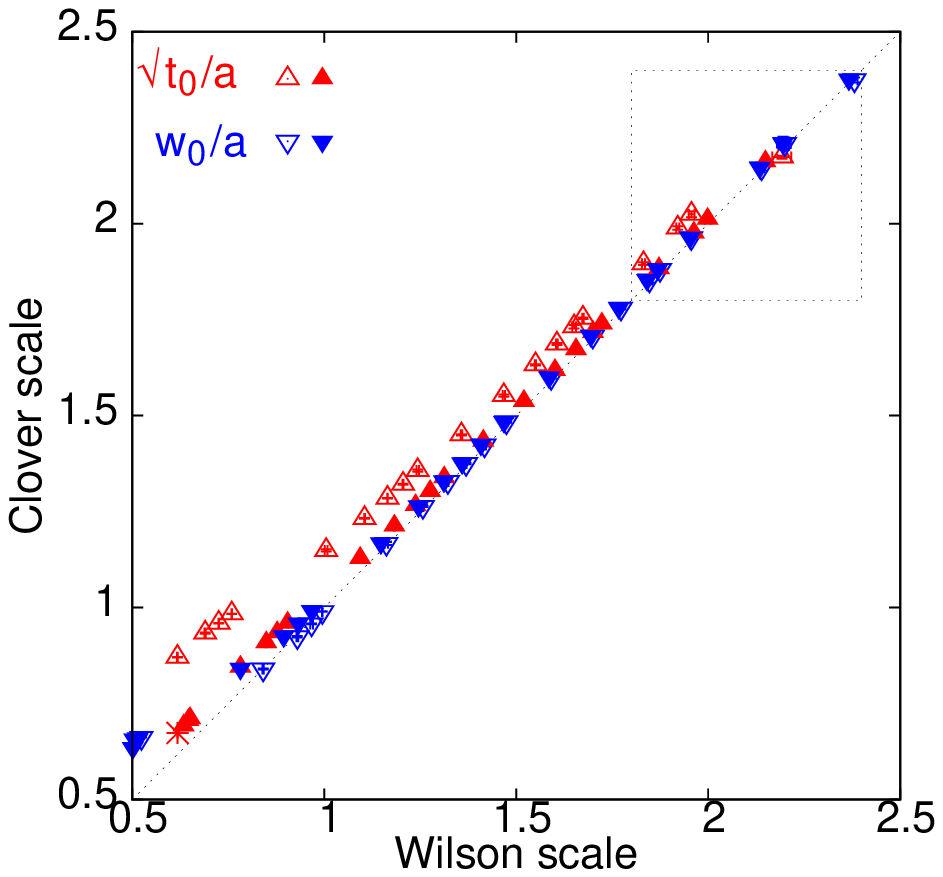}
\includegraphics[scale=0.7]{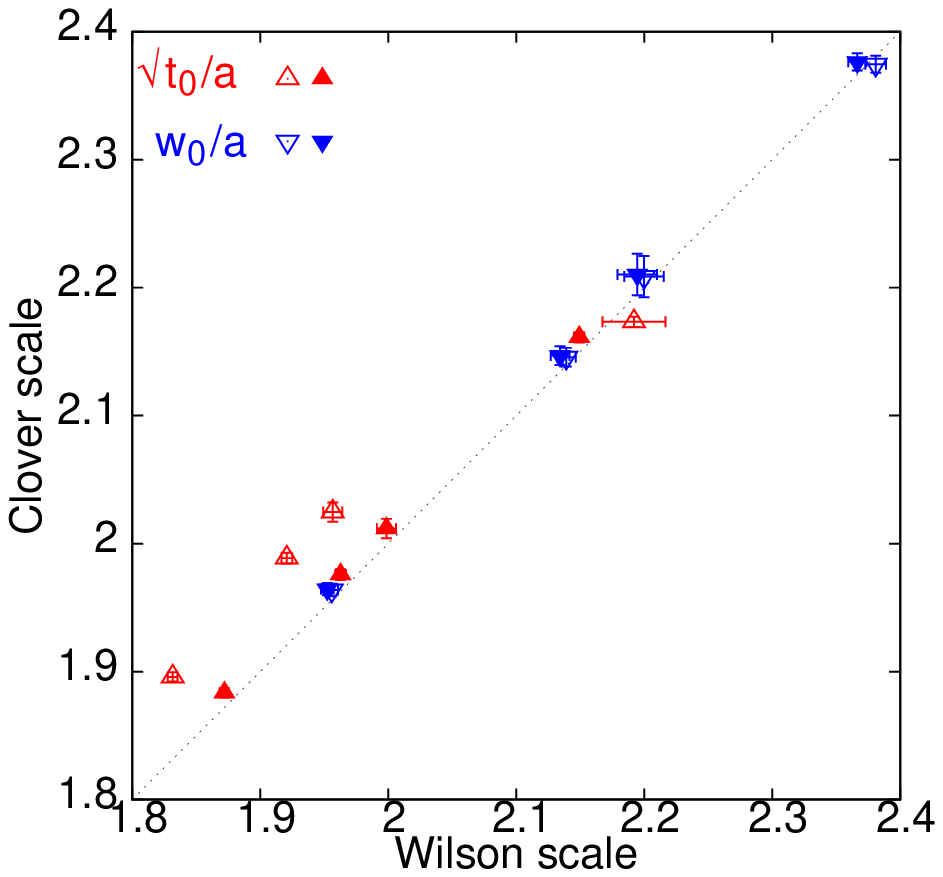}
\caption{The comparison of Wilson flow scales determined using the Wilson
 (plaquette) or the clover operator. The scale is varied by changing both
 gauge coupling and the bare quark mass. Up triangles denote $\sqrt t_0/a$,
 down triangles are for $w_0/a$. The unfilled symbols show the direct
 measurements, and filled symbols the tree-level improved values. The panel
 on the left collects all our measurements, that on the right shows only the
 four smallest lattice spacings, by zooming into the dotted box shown at
 the left.}
\eef{resulta}

\bef
\includegraphics[scale=1.0]{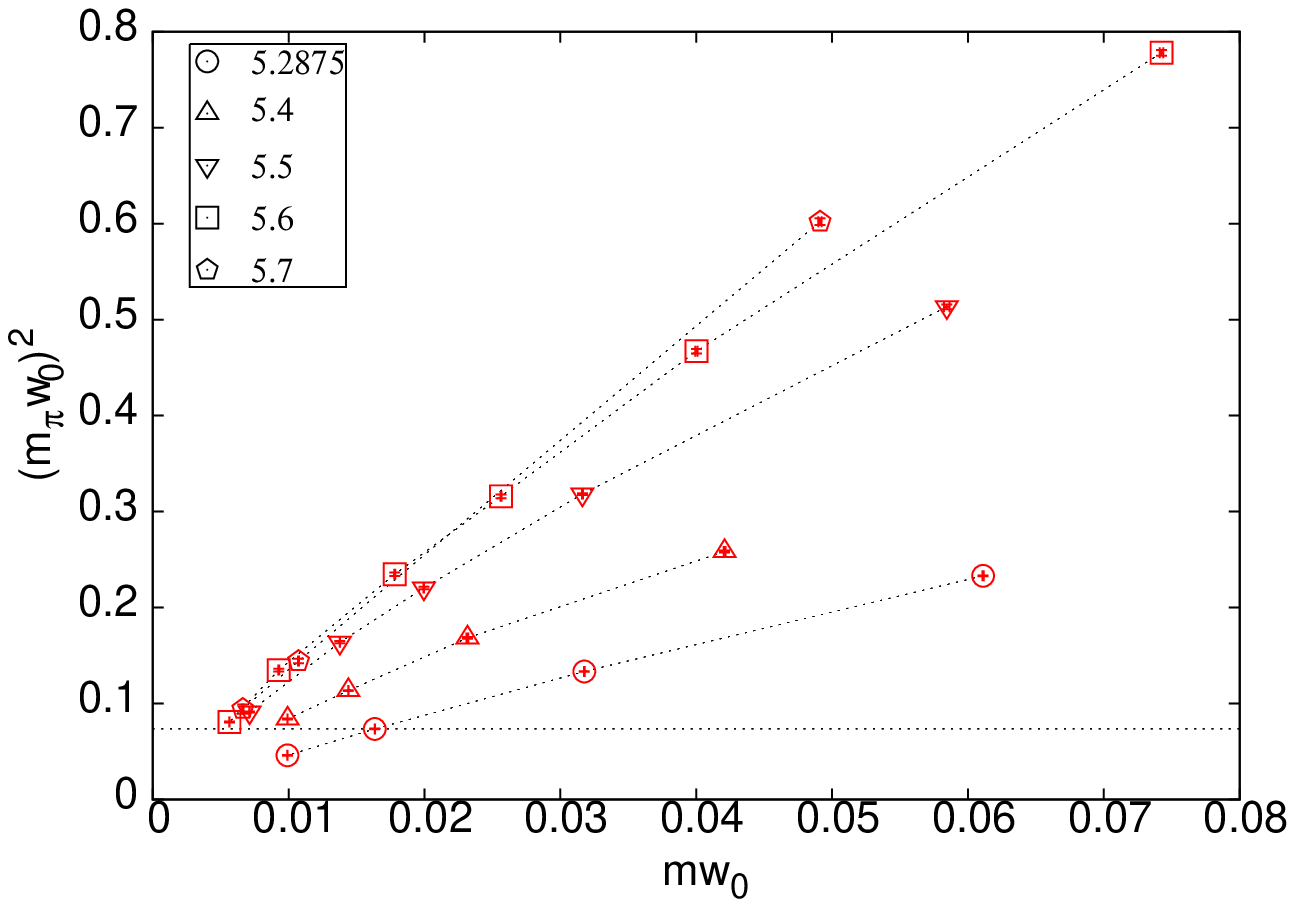}
\caption{The pion mass in physical units as a function of the bare quark 
 mass.}
\eef{massrenorm}

In \fgn{resulta} we plot a flow scale obtained with the Wilson
operator used for ${\cal E}(t)$ against the same scale obtained with the
clover operator. If the two were equal, then the measurements would
lie on the diagonal line. It has been observed before that the clover
improvement changes the flow scale $\sqrt{t_0}$ quite significantly,
as we verify again. The data set for the scale $w_0$ is significantly
closer to the diagonal. Both of these scales are improved significantly
by a tree-level improvement, at least on coarser lattice spacings: both
sets of measurements are moved significantly closer to the diagonal
line.  However, as shown in the zoom in \fgn{resulta}, the improvement
is marginal for $w_0$ at the smallest lattice spacings. This implies
that any remaining finite lattice spacing corrections in $w_0$ are
small. In view of this, we will use the tree-level improved value of $w_0$
to set the scale in the rest of the paper. We see that the range of lattice
spacings we scan covers a factor of four from the coarsest to the finest.

We apply this scale setting first to re-examine the pion mass
measurement.  Our measurements of $m_\pi$ in lattice units are given in
\tbn{configs}. We plot $m_\pi$ in units of $w_0$ in \fgn{massrenorm}. It
is clear from the figure that the range of pion masses explored in this
study covers a factor of four from the largest to the smallest.  Given the
rapid variation of $a/w_0$ and $m_\pi w_0$ with the bare coupling and the
bare quark mass, it is useful to trade the bare parameters for these two.

\bef
\includegraphics[scale=1.0]{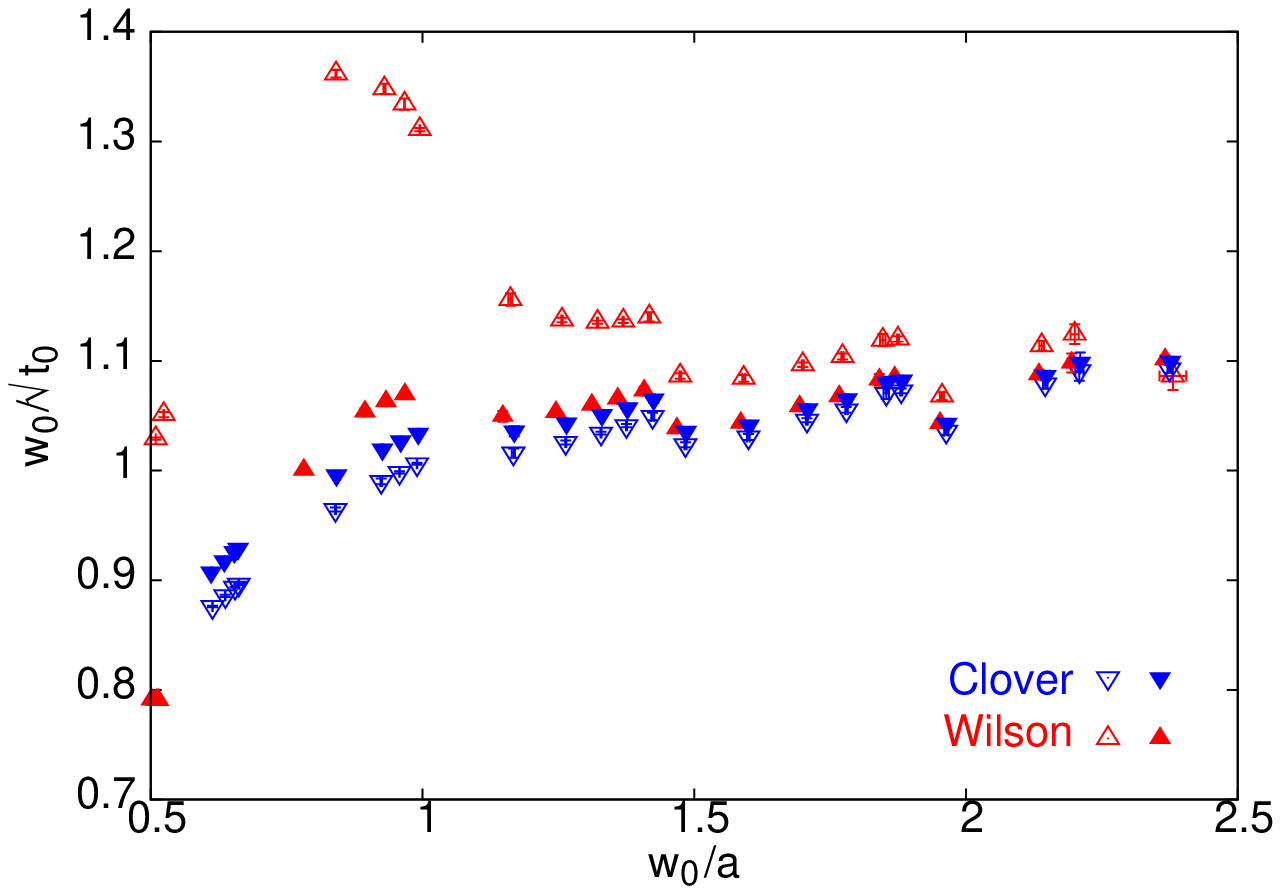}
\caption{The ratio of the scales $w_0/\sqrt{t_0}$ for all the bare couplings
 and quark masses studied. At the smallest lattice spacings the ratio
 from Wilson (up triangles) and clover (down triangles) operators, for both
 direct (unfilled symbols) and tree-level improved (filled symbols) scales
 tend to the same value.}
\eef{resultb}

\bef
\includegraphics[scale=0.6]{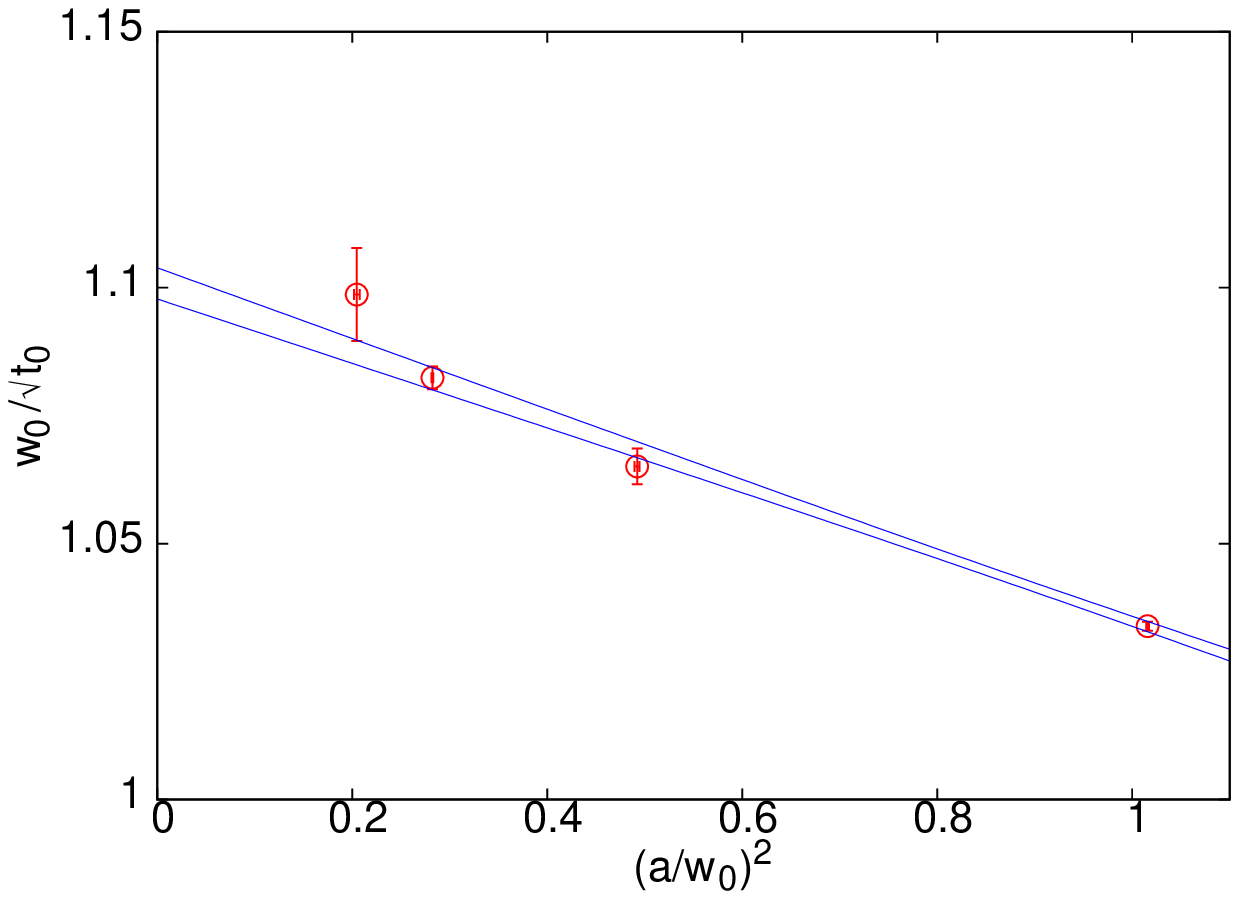}
\includegraphics[scale=0.6]{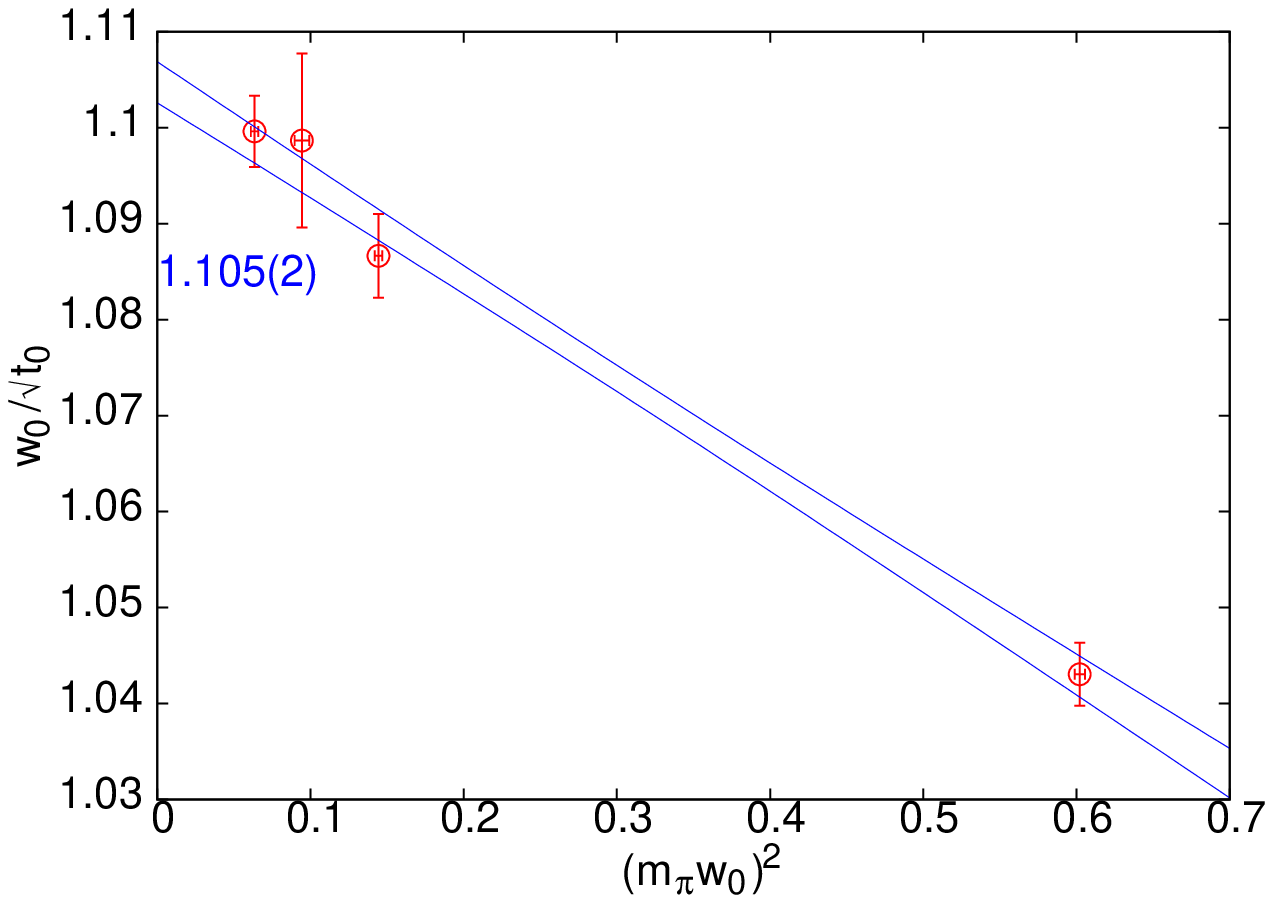}
\caption{(a) Variation of $w_0/\sqrt{t_0}$ with the lattice spacing at fixed
 $m_\pi w_0\simeq0.3$. The systematics of the continuum extrapolation is
 discussed in the text. (b) Variation of $w_0/\sqrt{t_0}$ with $m_\pi w_0$ at
 the smallest bare coupling. Chiral logs \cite{chirallog} are not visible
 at this precision.}
\eef{resultbx}

Since both the scales $\sqrt{t_0}$ and $w_0$ are physical, the ratio
${\cal R}=w_0 / \sqrt{t_0}$ is expected to tend to a good limit as the
lattice spacing decreases.  In \fgn{resultb} we show the dependence of
this ratio on the lattice spacing (given in units of the tree-level
corrected value of $w_0$). At the smallest lattice spacing which we
have examined ($w_0/a\simeq2.4$), ${\cal R}\simeq1.100 \pm 0.003$.
For 2+1+1 flavours of staggered quarks \cite{milcscale} we deduce ${\cal
R}\simeq1.21 \pm 0.01$, where the error is estimated conservatively
by neglecting covariance of the numerator and denominator. Since
the statistical errors in $\cal R$ are small, the difference is
significant. In a direct computation we checked that in the pure gauge
theory, when $w_0/a\simeq 2.4$, the ratio ${\cal R}=1.012 \pm 0.005$
(this is consistent with results presented in \cite{flowqcdscale}). The
ratio clearly depends on the number of flavours of quarks.

$\cal R$ also depends on the lattice spacing and the quark mass, as shown
in \fgn{resultbx}. At fixed renormalized quark mass, $m_\pi w_0\simeq0.3$
we have tried a quadratic extrapolation to the continuum. Using the data
points on the four finest lattices, the continuum extrapolated ratio is
$1.101 \pm 0.003$. A fit using the quartic term gives the extrapolated
value $1.11\pm0.01$. If one uses only the three finest lattices, then
the continuum extrapolation gives ${\cal R}=1.108\pm0.007$. We put these
observations together and quote a continuum extrapolated value
\beq
   {\cal R}=\frac{w_0}{\sqrt{t_0}} = \rval.
\eeq{continuum}
Following
\cite{sommer}, we define a measure of the slope with respect to the
lattice spacing as
\beq
   S^a_{\cal R} = \frac{{\cal R}(a=w_0/1.75)}{{\cal R}(a=0)} -1
   \simeq 14\%.
\eeq{aslope}
This is significantly larger than the results which can be reconstructed
from values for other slopes quoted for $N_f=2$ clover improved Wilson
fermions in \cite{sommer}. At this time we are unable to comment on what
combination of factors most influences this difference: the nature of
the sea quarks, the value of $m_\pi$, or technical issues in comparing
slopes of slightly different quantities \cite{sommer}.

\bef
\includegraphics[scale=1.0]{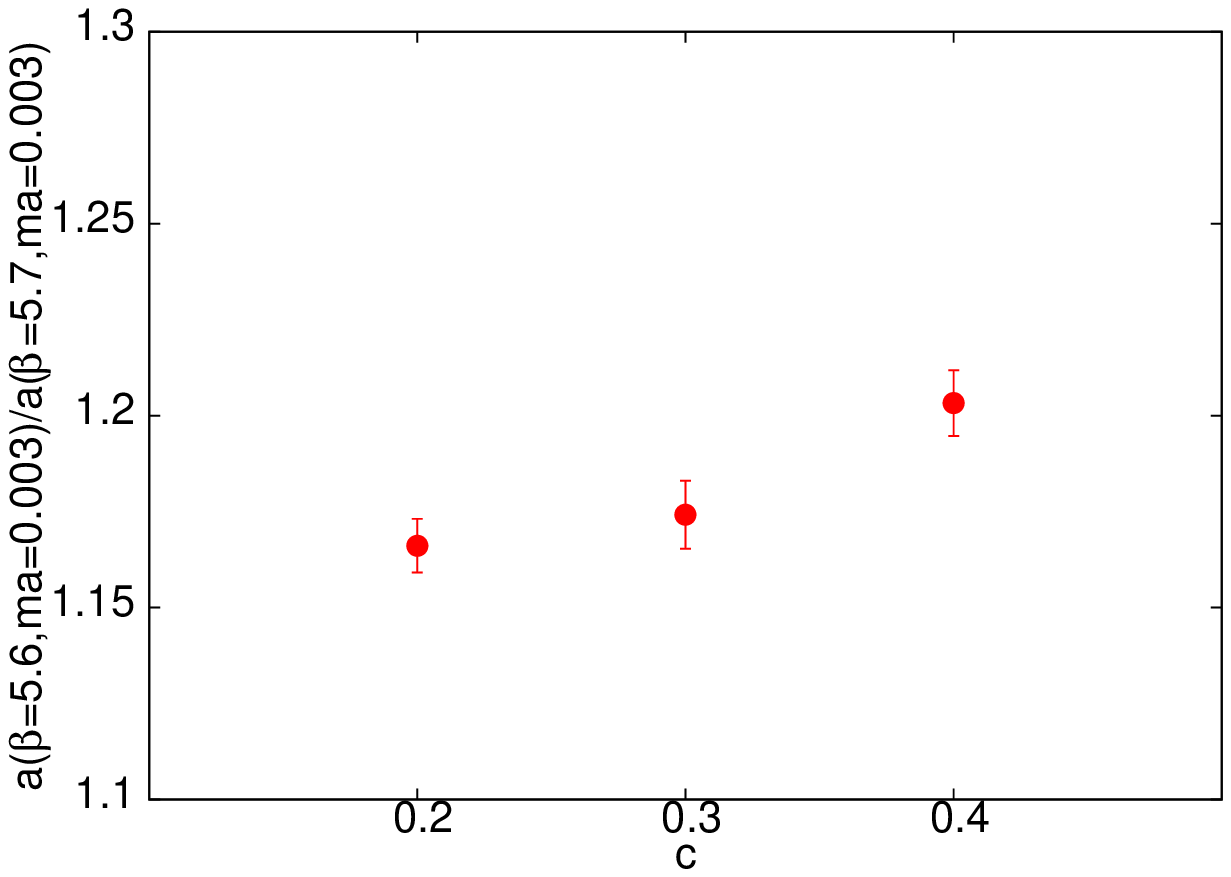}
\caption{The ratio of the lattice spacings for two different sets of
 bare parameters depends on the renormalization scheme, \ie, the
 choice of $c$, through $w_0(c)$.}
\eef{sensitivity}

\bef
\includegraphics[scale=0.6]{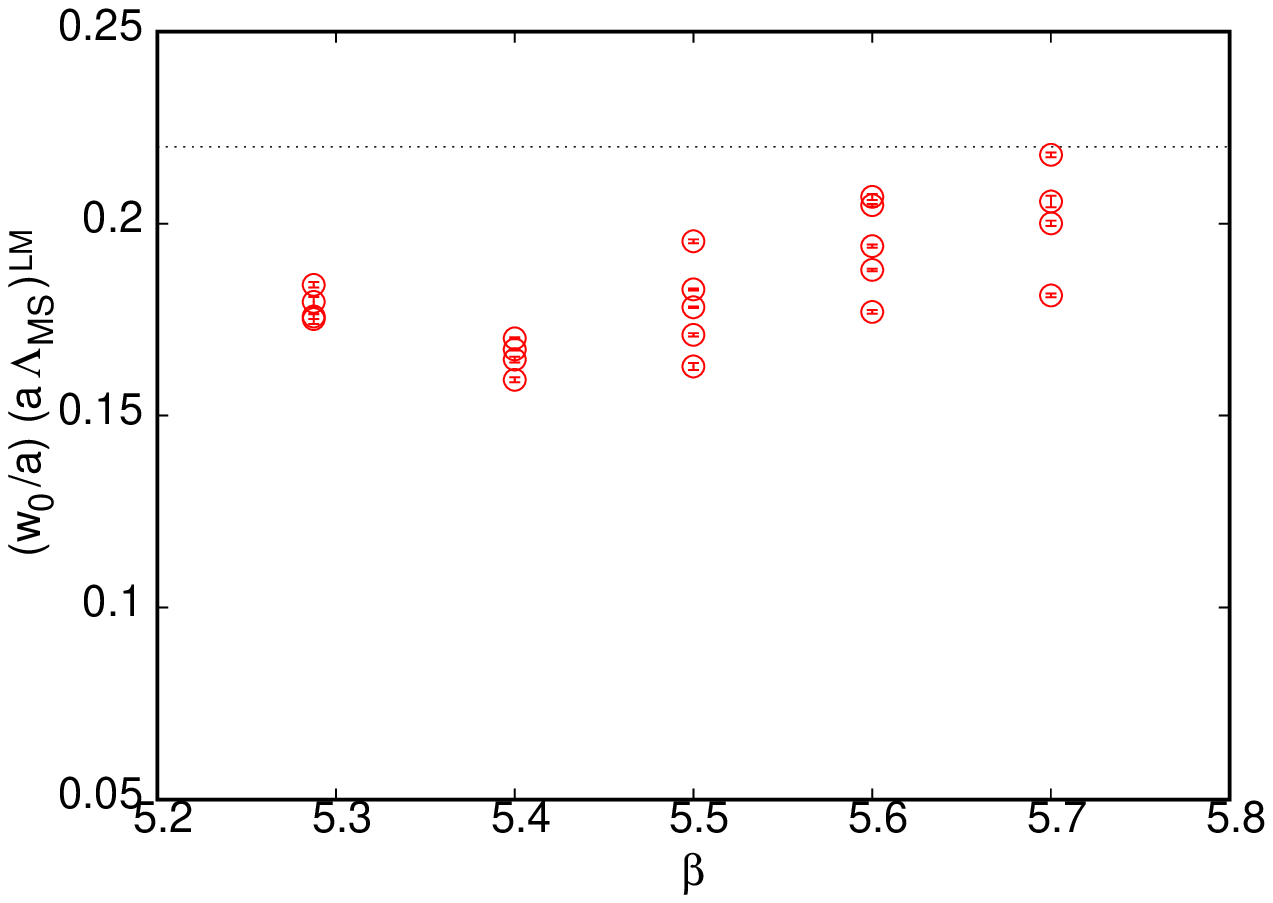}
\includegraphics[scale=0.6]{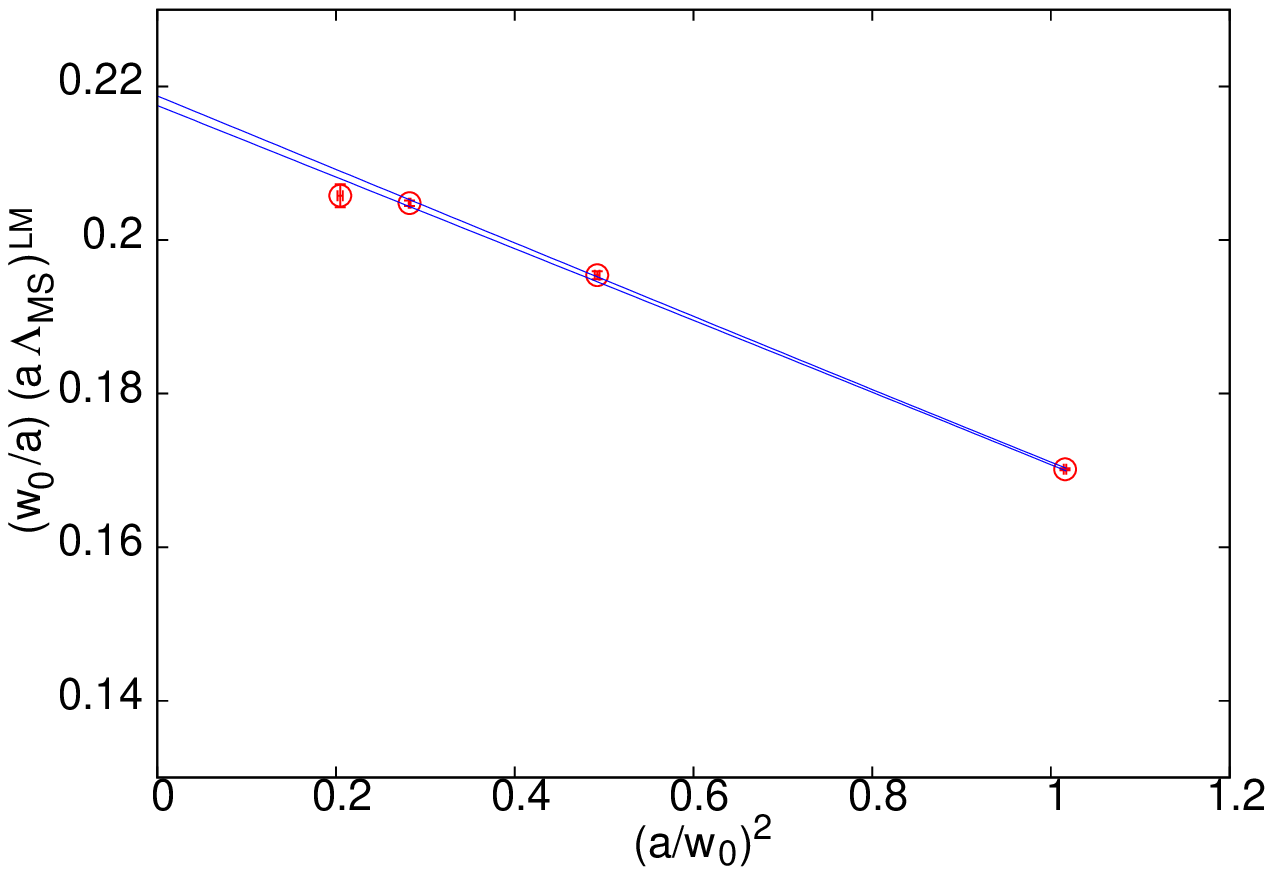}
\caption{The ratio $w_0 \Lambda_\MSbar$ extracted using the Lepage-Mackenzie
 scheme for computing $(a\Lambda_\MSbar)^{LM}$. The first panel shows
 all our simulation results plotted as a function of the bare coupling
 $\beta$. The next panel shows the results for fixed $m_\pi w_0\simeq0.3$
 and a quadratic continuum extrapolation.}
\eef{lepage}

It is known that $w_0$ is more strongly dependent on the quark mass than
$t_0$ \cite{sommer}.  A roughly linear dependence of both the scales
with the renormalized quark mass has been observed before over a range
of $m_\pi w_0$ similar to that explored here. \fgn{resultbx} shows this
linear behaviour of the ratio $w_0/\sqrt{t_0}$. An extrapolation to
the chiral limit as $m_\pi^2 w_0^2$ \footnote{Presumably when this
extrapolation is examined at smaller pion masses the subleading
corrections from chiral logs \cite{chirallog} will begin to be
numerically significant.} at our smallest bare coupling yields
$w_0/\sqrt{t_0}=1.105\pm0.002$ Using the $N_f=2$ value for $w_0$
above. Defining an effective slope parameter
\beq
  S^m_{\cal R} = \frac{{\cal R}(m_\pi w_0=0.45)-{\cal R}(m_\pi w_0=0.30)}
        {{\cal R}(m_\pi w_0=0.45)+{\cal R}(m_\pi w_0=0.30)}\,,
\eeq{chiralslope}
our observations give $S^m_{\cal R}\simeq2$\%.  This is compatible with
the change reported with two flavours of clover improved Wilson quarks
in \cite{sommer}.

Since the parameter $c$ determines the value of the running coupling
\eqn{coupling}, one may use the RG-flow of the coupling to examine the
$c$-dependence of $w_0(c)$.  Define a measure of the change in $w_0(c)$
through
\beq
  S_c = \frac{w_0(c=0.4)/a-w_0(c=0.2)/a}{2w_0(c=0.3)/a},
    \qquad (a{\rm\ fixed\/}).
\eeq{slopec}
On our finest lattice, we find $S_c\simeq0.1$. The same measure with
$t_0$ gives about 0.2.  The formal two-loop expression for the running of
$g_R$ in \eqn{coupling} yields $S_c \simeq 0.3$.  Since the renormalized
couplings obtained for these $c$ are large, the two-loop beta function
does not run the coupling reliably, so one should take the last number
only as indicating that such large changes in scale are natural when
changing $c$.

It is more interesting to ask whether the ratio of lattice spacings at two
different bare couplings and quark masses is independent of the choice of
$c$, when each of these is given in units of $w_0$.  Ideally, of course,
such a ratio should not change with $c$.  In \fgn{sensitivity} we show
that, in fact, there is some residual dependence on the parameter
$c$. Take the case where this ratio is close to 1.175 for $c=0.3$. The
change in this ratio of lattice spacings for variation of $\Delta c=0.2$
around $c=0.3$ is 3\% of the central value.  While not ideal, this change
is rather small. Presumably this uncertainty in the scale setting is
due to remaining lattice spacing corrections. It would be interesting
in the future to perform this comparison at smaller lattice spacings.

Measurements of plaquettes can also be converted to a scale using the
methods of \cite{lm}. Since the scale setting by the flowed plaquette
suffers from significant lattice spacing effects at flow times $w_0^2$,
necessitating the various corrections which we have explored, it may be
suspected that these effects could be larger at flow time $t=0$. These
are partly taken into account by corrections suggested in \cite{ehk}.
In \fgn{lepage} we show the dimensionless ratio $w_0\Lambda_\MSbar$
obtained by a comparison of this scale with the flow scale $w_0/a$. In
the second panel of \fgn{lepage}, we show the ratio $w_0\Lambda_\MSbar$
at fixed pion mass, $m_\pi w_0\simeq0.3$ as a function of the lattice
spacing. One sees a strong, nearly quadratic, lattice spacing dependence,
albeit with a slope smaller than $S^a_{\cal R}$.  A quadratic extrapolation
to the continuum limit gives $w_0\Lambda_\MSbar=0.218\pm0.001$, where the
error is statistical only.  It is interesting to compare this indicative
number to the value for $N_f=2$ clover fermions. We take $\Lambda_\MSbar
= 257 \pm 26$ MeV as quoted in \cite{alpha}, and combine it with the
value of $w_0$ reported with $N_f=2$ clover fermions \cite{sommer},
to get $w_0\Lambda_\MSbar=0.23\pm0.03$.

\section{Conclusions}\label{sec:conclude}

We have reported on investigations of the Wilson flow scales $\sqrt{t_0}$
and $w_0$ in QCD with two flavours of naive staggered quarks. Our
investigations cover a wide range of lattice spacings (a factor of about
4) and pion masses (also a factor of about 4). We found that the scale
$w_0$ has smaller lattice spacing artifacts than $\sqrt{t_0}$. One
consequence of this is that tree-level improvement of the former has
smaller effect than in the latter. In most of this paper we have used
tree-level improved measurements of $w_0$ obtained from the clover
operator as the object to set the scale by.

We found an interesting approximate scaling of the autocorrelations of the
basic measurement $\langle{\cal E}(t)\rangle$. The integrated autocorrelation
time increases with $t$ before saturating. The scaling implies that keeping
the error in measurements of $w_0$ fixed in the continuum limit may require
the statistics to grow as the inverse square of the lattice spacing.

We found that the ratio ${\cal R}=w_0/\sqrt{t_0}=1.100\pm0.003$ when
$w_0/a\simeq2.4$. A continuum extrapolation at fixed $m_\pi w_0\simeq0.3$
gave ${\cal R}=\rval$. Comparison with results for the pure gauge theory,
and with $N_f=2+1+1$ reveals a dependence of $\cal R$ on $N_f$. The the
compilation of \cite{sommer} also shows this trend for staggered quarks,
but not for Wilson quarks. For $N_f=2$ clover improved Wilson quarks,
the value of $\cal R$ is different from our determination \cite{sommer}.

The dependence of the scale $w_0(c)$ on $c$ is large; this is natural
since $c$ enters linearly in the definition of $g_R^2$, which depends
nearly logarithmically on the scale $w_0(c)$. In principle, this should
not change the ratio of two lattice spacings. However, we found a mild
(3\%) dependence of the ratio of two lattice spacings on $c$.  The effect
is small enough that one suspects it is due to lattice spacing dependences
which are not absorbed into the tree-level improvement of $w_0$.

By using our data sets to determine the scale via the Lepage-Mackenzie
prescription \cite{lm} we found that it has large lattice
spacing corrections.  However, with our data we tried a simple
continuum extrapolation at fixed $m_\pi w_0\simeq0.3$, and found
$w_0\Lambda_\MSbar=0.218 \pm 0.001$. If one then uses the ALPHA
collaboration's value $\Lambda_\MSbar=257\pm26$ MeV for $N_f=2$, then one
is led to the conclusion that for naive staggered quarks $w_0=0.17\pm0.02$
fm. This error is purely statistical, and dominated by the statistical
error in the determination of $\Lambda_\MSbar$.

{\bf Acknowledgements}: These computations were performed with the Cray
X1 and IBM Blue Gene/P installations of the ILGTI in Mumbai, and with
the Cray XK6 installation of the ILGTI in Kolkata.


\begin{thebibliography}{99}
\bibitem{flow}
 M.\ L\"uscher, \jhep 1008 (2010) 071 [arxiv:1006.4518], \jhep 1403 (2014) 092;
 M.\ L\"uscher, arxiv:1101.0962,
\bibitem{also}
 R.\ Narayanan and H.\ Neuberger, \jhep 0603 (2006) 064 [hep-th/0601210];
 R.\ Lohmayer and H.\ Neuberger, {\sl PoS\/} LATTICE2011 (2011) 249 [arxiv:1110.3522].
\bibitem{bmw}
 S.\ Borsanyi \etal, \jhep 1209 (2012) 010 [arxiv:1203.4469].
\bibitem{sommer}
 R.\ Sommer, {\sl PoS\/} LATTICE2013 (2014) 015 [arxiv:1401.3270];
 R.\ Sommer and U.\ Wolff, {\sl Nucl.\ Part.\ Phys.\ Proc.\/} 261--262 (2015) 155 [arxiv:1501.01861].
\bibitem{flowqcdscale}
 A.\ Francis \etal, \prd 91 (2015) 096002 [arxiv:1503.05652];
 M.\ Asakawa \etal, arxiv:1503.06516.
\bibitem{wilson2}
 M.\ Bruno and R.\ Sommer, {\sl PoS\/} LATTICE2013 (2014) 321 [arxiv:1311.5585].
\bibitem{wilson21}
 R.\ Horsley \etal, {\sl PoS\/} LATTICE2013 (2014) 249 [arxiv:1311.5010].
\bibitem{milcscale}
 R.\ J.\ Dowdall \etal, \prd 88 (2013) 074504 [arxiv:1303.1670];
 A. Bazavov \etal (MILC) arxiv:1503.02769.
\bibitem{montvay}
 I.\ Montvay and G.\ Muenster, {\sl Quantum Fields on a Lattice\/},
 Cambridge University Press (1994).
\bibitem{tlimprov}
 Z.\ Fodor \etal, \jhep 1409 (2014) 018 [arxiv:1406.0827].
\bibitem{volzero}
 Z.\ Fodor \etal, \jhep 1211 (2012) 007 [arxiv:1208.1051].
\bibitem{nf2}
 S.\ Gottlieb \etal, \prd 38 (1988) 2235;
 K.\ M.\ Bitar \etal, \prd 42 (1990) 3794;
 F.\ R.\ Brown \etal, \prl 67 (1991) 1062.
\bibitem{hmc}
 S.\ Gupta, \np B 370 (1992) 741.
\bibitem{chirallog}
 O.\ B\"ar and M.\ Golterman, \prd 89 (2014) 034505 [arxiv:1312.4999].
\bibitem{lm}
 G.\ P.\ Lepage and P.\ B.\ Mackenzie, \prd 48 (1993) 2250 [hep-lat/9209022].
\bibitem{ehk}
 R.\ G.\ Edwards \etal, \np B 517 (1998) 377.
\bibitem{alpha}
 M.\ Della Morte \etal, \jhep 0807 (2008) 037.
\end{thebibliography}
\end{document}